\documentclass{jpsj-suppl}
\usepackage{txfonts} 
\def\slashchar#1{\setbox0=\hbox{$#1$}
   \dimen0=\wd0 \setbox1=\hbox{/} \dimen1=\wd1
   \ifdim\dimen0>\dimen1 \rlap{\hbox to \dimen0{\hfil/\hfil}} #1
   \else  \rlap{\hbox to \dimen1{\hfil$#1$\hfil}} / \fi}

\title{Pion production off the nucleon}

\author{M. Rafi  \textsc{Alam}, M. Sajjad \textsc{Athar}, Shikha  \textsc{Chauhan}  and
S. K. \textsc{Singh}}

\inst{Department of Physics, Aligarh Muslim University, Aligarh, India}

\email{rafi.alam.amu@gmail.com}

\recdate{\today}
\abst{We have studied charged current neutrino/antineutrino induced weak pion production from nucleon.
For the present study, contributions from $\Delta(1232)$-resonant term, non-resonant background terms as well as 
contribution from higher resonances viz. $P_{11}$(1440), $D_{13}$(1520), $S_{11}$(1535), $S_{11}$(1650) and $P_{13}$(1720) are taken.
 To write the hadronic current for the non-resonant background terms,  
 a microscopic approach based on SU(2) non-linear sigma model has been used.
       The vector form factors for the resonances are obtained from  
     the helicity amplitudes provided by MAID. 
     Axial coupling in the case of $\Delta(1232)$ resonance is obtained by fitting 
     the ANL and BNL $\nu$-deuteron reanalyzed scattering  data. 
     The results of the cross sections are presented and discussed for all the possible 
     channels of single pion production induced by charged current interaction.}
\kword{Pion production, Deuteron effects, Higher resonances, Charged current.}

\begin{document}
\maketitle

\section{Introduction}
Experimenters are using neutrino/antineutrino beam of few GeV energy in the study of neutrino/antineutrino nucleus scattering  
 to determine some of the oscillation parameters like $\Delta m_{32}^2$, $\theta_{32}$, 
 CP violating phase $\delta$, etc.
 They are also important because of the interest in understanding hadronic structure in 
 weak sector where besides vector current they also get contribution from axial vector current. 
 In the neutrino/antineutrino energy region of $~$1GeV, the dominant contribution to the charged lepton  events come from 
 quasielastic scattering and single pion production processes. 
 The available experimental results of single pion production  and their comparison with various 
theoretical calculations have necessitated the need to re-examine the basic reaction mechanism for the production of single pion
from free nucleon target. In various theoretical calculations there is lack of consensus in the modeling of basic reaction 
mechanism of $\nu(\bar\nu)$ induced pion production 
 from free nucleon, specially concerning the contribution of background terms as well as the 
 contribution of higher resonances in addition to the 
 dominant $\Delta$(1232) resonance. The tension between the experimental results from old bubble chamber 
 experiments, ANL~\cite{Radecky:1981fn} and BNL~\cite{Kitagaki:1986ct}, 
  which were performed using deuterium/hydrogen targets, necessitated the need of re-performing the
  experiments with high precision using deuterium targets.  Recently, Wilkinson et al.~\cite{Wilkinson:2014yfa} have reanalyzed the old 
 ANL~\cite{Radecky:1981fn} and BNL~\cite{Kitagaki:1986ct} data 
 and found the differences in these two results to be within 12$\%$ at $E_\nu = 1 GeV$ and 
 8$\%$ at $E_\nu = 2 GeV$. 
 A theoretical understanding of these data may be of great interest in the understanding of hadronic physics.
  
     In this work, we present a study of single pion production induced by neutrinos/antineutrinos off nucleon. 
 Besides the dominant $\Delta(1232)$-term, we have considered 
 non-resonant background terms and also taken the contribution of higher resonances 
 viz. $P_{11}$(1440), $D_{13}$(1520), $S_{11}$(1535), $S_{11}$(1650) and $P_{13}$(1720). 
  Presently there is no consensus as to how the non-resonant background terms 
 should be added to the dominant $\Delta(1232)$ contribution. Some authors have performed 
 calculations by coherently summing the contributions of hadronic current 
 from the background terms and the $\Delta(1232)$-resonant term, while some have added them incoherently.
 The understanding of the role of background terms is specially important in determining the $N-\Delta$ 
transition form factors in $  \nu_\mu p \to \mu^- p \pi^+ $ and $\bar \nu_\mu n \to \mu^+ n \pi^-$ channels 
which are dominated by $\Delta(1232)$- excitation and receive no contribution from the nearby higher resonance which are $I = \frac12$ resonances.

In section-\ref{sec:nr_back}, we present the formalism in brief and discuss the results in section-\ref{results}.  
\section{Formalism}\label{sec:nr_back}
The cross section for the single pion production process $\nu_l(\bar\nu_l)~+~N~\rightarrow~l^-
(l^+)~+~N^\prime~+~\pi^i;~~N,N^\prime=p,n; i=\pm,0$, may be written as, 
\begin{align}\label{eq:phsp}
 d\sigma = \frac{(2\pi)^{4}}{4 ME} \delta^{4}(k+p-k^{\prime}-p^{\prime}-k_{\pi}) 
\frac{d{\vec p\,}^{\prime}}{ (2\pi)^{3}  2 E^{\prime}_{p} } \frac{d{\vec k}_{\pi}}{(2\pi)^{3}  2 E_{\pi} }
  \frac{d{\vec k}^{\prime}}{  (2\pi)^{3} 2 E_{l} } \bar\Sigma\Sigma | \mathcal M |^2
\end{align}
where $ k( k^\prime) $ is the four momentum of the incoming(outgoing) lepton having energy $E(E_l)$ 
while   $p( p^\prime)$ is the four momentum of the incoming(outgoing)
nucleon and the pion momentum is $k_\pi $ having energy $ E_\pi  $.
 $| \mathcal M |^2$ is the square of the matrix element given by
\begin{align}
  | \mathcal M |^2  = \frac{G_F^2}{2}\, L_{\mu \nu} J^{\mu \nu}.
\end{align}
where $L_{\mu \nu}$ is the leptonic tensor 
\begin{align}
L_{\mu \nu} = l_\mu l_\nu^\dagger = 8 \left( k_\mu k_\nu^\prime+k_\mu^\prime k_\nu-g_{\mu\nu}~k\cdot k^\prime 
\pm i \epsilon_{\mu\nu\alpha\beta}~k^{\prime \alpha} k^{\beta}\right), 
\end{align}
the upper(lower) sign in the antisymmetric term stands for (antineutrino)neutrino induced processes.

\begin{figure}[tbh]
\centerline{\includegraphics[height=9cm]{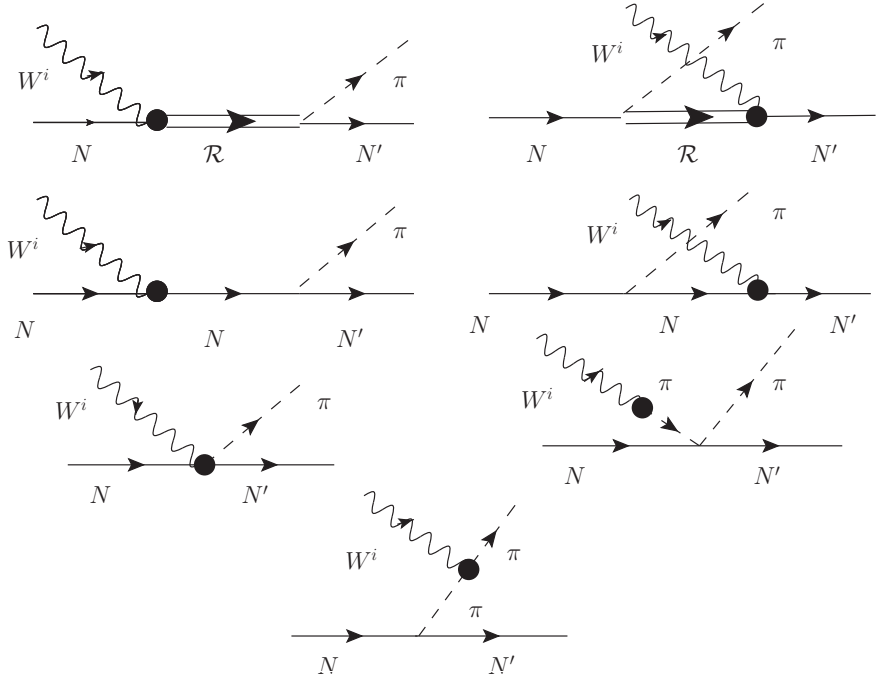}}
\caption{Feynman diagrams contributing to the hadronic current corresponding 
to $W^{i} N \to N^{\prime} \pi^{\pm,0}$, where $(W^i \equiv W^\pm \; ; i=\pm)$ for charged 
current processes with $N,N^{\prime}=p \;{\rm or}\; n$. First row: direct and cross diagrams for resonance 
production where intermediate term $R$ stands for different resonances. Second row: nucleon pole(NP and CNP) 
terms. The third row the diagrams are for contact term(CT) and pion pole(PP)
term (third row left to right) and pion in flight(PF)(fourth row)  terms.}
\label{fig:feyn}
\end{figure}

To get the expression for hadronic tensor $J^{\mu \nu}(= j_\mu j_\nu^\dagger) $, the hadronic current
has been obtained for the Feynman diagrams shown in  Fig~\ref{fig:feyn}.
 The contributions of non-resonant background terms are obtained using a chiral invariant Lagrangian based on 
non-linear sigma model~\cite{Hernandez:2010bx}. 
At tree level the various diagrams which may contribute to the pion-production mechanism are 
direct and cross nucleon pole, contact diagram, pion pole and pion in flight diagram 
labeled  as NP, CNP, CT, PP and  PF respectively.
As the non-linear sigma model assumes hadrons as point like particle, therefore, to take into account the structure of hadron, 
the form factors are introduced at the $W^{\pm} N \rightarrow N$ transition vertex. 
The details are given in Refs.~\cite{Hernandez:2007qq,aip1}. 

The final form of hadronic current for non-resonant background are obtained as~\cite{Hernandez:2007qq}
\begin{eqnarray} \label{eq:background}
j^\mu\big|_{NP} &=& 
\mathcal{A}^{NP}
  \bar u(\vec{p}\,') 
 \slashchar{k}_\pi\gamma_5\frac{\slashchar{p}+\slashchar{q}+M}{(p+q)^2-M^2+ i\epsilon}\left [V^\mu_N(q)-A^\mu_N(q) \right]  
u(\vec{p}\,),\nonumber \\\nonumber \\
j^\mu\big|_{CP} &=& 
\mathcal{A}^{CP}
  \bar u(\vec{p}\,') \left [V^\mu_N(q)-A^\mu_N(q) \right]
\frac{\slashchar{p}'-\slashchar{q}+M}{(p'-q)^2-M^2+ i\epsilon} 
\slashchar{k}_\pi\gamma_5  u(\vec{p}\,),\nonumber \\\nonumber \\
j^\mu\big|_{CT} &=&
\mathcal{A}^{CT}
  \bar u(\vec{p}\,') \gamma^\mu\left (
  g_A  f_{CT}^V(Q^2)\gamma_5 - f_\rho\left((q-k_\pi)^2\right) \right ) u(\vec{p}\,),\nonumber \\\nonumber \\
j^\mu\big|_{PP} &=& 
\mathcal{A}^{PP}f_\rho\left((q-k_\pi)^2\right)
  \frac{q^\mu}{m_\pi^2+Q^2}
  \bar u(\vec{p}\,')\ \slashchar{q} \ u(\vec{p}\,),\nonumber \\ \nonumber \\
j^\mu\big|_{PF} &=& 
\mathcal{A}^{PF}f_{PF}(Q^2)
  \frac{(2k_\pi-q)^\mu}{(k_\pi-q)^2-m_\pi^2}
  2M\bar u(\vec{p}\,')  \gamma_5 u(\vec{p}\,),\label{eq:eqscc}
\end{eqnarray}
where for NP and CNP currents, at the  transition vertex $W^\pm N \to N$, 
we have introduced form factors in $V^\mu_N(q)$ and $A^\mu_N(q)$ to account for the nucleon structure,
given by
\begin{eqnarray}\label{eq:vec_curr}
V^{\mu, CC}_N(q)&=&f_1(Q^2)\gamma^\mu + f_2(Q^2)i\sigma^{\mu\nu}\frac{q_\nu}{2M} \\
\label{eq:axi_curr}
A^{\mu, CC}_N(q)&=& \left(f_A(Q^2)\gamma^\mu 
+ f_P(Q^2) \frac{q^\mu}{M}  \right)\gamma^5,
\end{eqnarray}
where  $f_{1,2}(Q^2)$ and $f_{A,P}(Q^2)$ are the isovector vector and axial vector form factors for nucleons. 
Similarly $f_{\rho}(Q^2)$ accounts for dominant contribution that comes from $\rho$--meson cloud at 
$\pi \pi NN$ vertex in the case of PP diagram. 
Finally,  CVC relates the $ f_{PF}(Q^2) , f_{CT}^{V}(Q^2)$ with $f_{1} (Q^2)$ and 
PCAC relates $f_{\rho}(Q^2)$ with the axial part of CT diagrams. 
In the case of resonances, apart from the dominant $\Delta(1232)$ resonance, 
we have also  considered the contribution of various resonances from  second resonance region viz:
$D_{13}(1520)$ and $P_{13}(1720)$  which have $J=\frac32 , \, I=\frac12  $
and $P_{11}(1440)$, $S_{11}(1535)$ and $S_{11}(1650)$ which have $J=\frac12 , \, I=\frac12  $. 
 The Feynman diagrams for the resonant contributions are depicted in Fig.~\ref{fig:feyn} and 
are labeled as $R$ and $CR$ corresponding to direct and cross terms. 
 $ R/CR$ represents both spin half and three half resonances. 
The currents for spin $J=\frac32$ resonances are obtained as 
\begin{eqnarray}\label{eq:res_had_current}
j^\mu\big|_{R}^{\frac32} &=& i \; \cos~\theta_c\;  \mathcal{C}^{R} 
   \frac{k_\pi^\alpha}{p_R^2-M_R^2+ i M_R \Gamma_R}
   \bar u(\vec{p}\,') {\mathcal P}_{\alpha\beta}^{3/2}(p_R) \Gamma^{\beta\mu}_{\frac32}(p,q)
   u(\vec{p}\,),\quad p_R=p+q, \nonumber  \\ 
 j^\mu\big|_{C R}^{\frac32} &=& i \; \cos~\theta_c\;  \mathcal{C}^{R} 
   \frac{k_\pi^\beta}{p_R^2-M_R^2+ i M_R \Gamma_R}
   \bar u(\vec{p}\,')  {\hat \Gamma}^{\mu\alpha}_{\frac32}(p',-q) {\mathcal P}_{\alpha\beta}^{3/2}(p_R) 
   u(\vec{p}\,),\quad p_R=p'-q,  
\end{eqnarray}
and for spin $J=\frac12$ resonances are obtained as
\begin{eqnarray}\label{eq:res1/2_had_current}
j^\mu\big|_{R}^{\frac12}&=& 
i \; \cos~\theta_c\;  \mathcal{C}^{R} 
  \bar u(\vec{p}\,') 
 \slashchar{k}_\pi\gamma_5\frac{\slashchar{p}+\slashchar{q}+M_R}{(p+q)^2-M_R^2+ i \Gamma_R M_R  }\Gamma^\mu_{\frac12}  
u(\vec{p}\,),\nonumber \\\nonumber \\
j^\mu\big|_{C R}^{\frac12} &=& 
i \; \cos~\theta_c\; \mathcal{C}^{R} 
  \bar u(\vec{p}\,') \Gamma^\mu_{\frac12}
\frac{\slashchar{p}'-\slashchar{q}+M_R}{(p'-q)^2-M_R^2+ i \Gamma_R M_R } \slashchar{k}_\pi\gamma_5  u(\vec{p}\,),
\end{eqnarray}
where $ \mathcal{C}^{R} $ is the coupling strength for $R \to N\pi$ and $M_R$ is the 
mass of the resonance. ${\mathcal P}_{\alpha\beta}^{3/2}$ 
is spin three-half projection operator and is given by
\begin{equation}
{\mathcal P}_{\alpha\beta}^{3/2}(P)=- \left(\slashchar P \, + M_R \right) \left( g_{\alpha \beta}
- \frac{2}{3} \frac{P_{\alpha } P_{\beta}}{M_R^2} 
+ \frac{1}{3} \frac{P_{\alpha } \gamma_{\beta} - P_{\beta } \gamma_{\alpha}}{M_R} 
- \frac{1}{3} \gamma_{\alpha} \gamma_{\beta} \right),
\end{equation}
\begin{table*}[]
  \begin{center}
    \vspace{1cm}
    \begin{tabular*}{139mm}{@{\extracolsep{\fill}}c c c c c c c c c}
      \noalign{\vspace{-8pt}}
      \hline \hline
      Resonances               & $M_R$ [GeV] & J\quad   & I \quad    &   P   & $\Gamma_0^{tot}$  &  $\pi N$ branching  & $F_A(0)$ &   $f^\star$  \\
      &&&&&(GeV)&ratio ($\%$)  & or  ${\tilde C}_5^A(0)$\\ \hline
      $P_{33}$(1232)  &       $1.232$ & $3/2$ &$ 3/2$ &$ +$ &                $ 0.120$ &       $ 100$ &$ 1.0$ &   $2.14 $    \\ \hline
      
      $P_{11}$(1440)  &       $1.462$ & $1/2$ &$ 1/2$ &$ +$ &                $ 0.250$ &       $ 65$ &$ -0.43$&   $0.215 $	 \\ \hline
      
      $D_{13}$(1520)  &       $1.524$ & $3/2$ &$ 1/2$ &$ -$ &                $ 0.110$ &       $ 60$ &$-2.08$&  $1.575 $	 \\ \hline
      
      $S_{11}$(1535)  &       $1.534$ & $1/2$ &$ 1/2$ &$ -$ &                $ 0.151$ &       $ 51$ &$0.184$&	   $ 0.092 $	 \\ \hline
      
      $S_{11}$(1650)  &       $1.659$ & $1/2$ &$ 1/2$ &$ -$ &                $ 0.173$ &        $ 89$ &$-0.21$& $ - 0.105 $	 \\ \hline
      
      $P_{13}$(1720)  &       $1.717$ & $3/2$ &$ 1/2$ &$ +$ &                $ 0.200$ &       $ 11$ &$ -0.195$& $ 0.147 $	  \\ \hline
      \hline \hline
    \end{tabular*}
  \end{center}
  \caption{Properties of the 
  resonances included in the present model, with Breit-Wigner mass $M_R$, spin J, 
  isospin I, parity P, the total decay width $\Gamma_0^{tot}$, 
  the branching ratio into $\pi$ N, the axial coupling and $f^\star$.\cite{aip1}}
   \label{tab:included-resonances}
\end{table*}
The weak vertex $\Gamma^{\nu\mu} (\Gamma^{\mu})  $ for spin $\frac32(\frac12)$ resonances has V-A structure, given by
\begin{align}\label{eq:vec_3half_pos}
  \Gamma_{\nu \mu }^{\frac{3}{2}^+} =& \left[ {V}_{\nu \mu }^\frac{3}{2} - {A}_{\nu \mu }^\frac{3}{2} \right] \gamma_5
  & \quad \quad &    \Gamma_{ \mu }^{\frac{1}{2}^+}  =   V_\mu^{\frac{1}{2}} - A_\mu^{\frac{1}{2}}      \nonumber\\
  \Gamma_{\nu \mu }^{\frac{3}{2}^-} =&   {V}_{\nu \mu }^\frac{3}{2} - {A}_{\nu \mu }^\frac{3}{2}  
   & \quad \quad &   \Gamma_{ \mu }^{\frac{1}{2}^-}  =  \left[ V_\mu^{\frac{1}{2}} - A_\mu^{\frac{1}{2}} \right] \gamma_5
\end{align}
where the superscript $+(-)$ stands for positive(negative) parity state.
For spin three half states the vertex $\Gamma_{\nu \mu }$ may be written in terms of six form factors viz:
\begin{eqnarray}\label{eq:vec_axial}
  V_{\nu \mu }^{\frac{3}{2}} &=&   \frac{{\tilde C}_3^V}{M} (g_{\mu \nu} \slashchar{q} \, - q_{\nu} \gamma_{\mu})+
  \frac{{\tilde C}_4^V}{M^2} (g_{\mu \nu} q\cdot p' - q_{\nu} p'_{\mu}) 
  + \frac{{\tilde C}_5^V}{M^2} (g_{\mu \nu} q\cdot p - q_{\nu} p_{\mu}) + 
  g_{\mu \nu} {\tilde C}_6^V   \nonumber\\
  A_{\nu \mu }^{\frac{3}{2}} &=&- \left[ \frac{{\tilde C}_3^A}{M} (g_{\mu \nu} \slashchar{q} \, - q_{\nu} \gamma_{\mu})+
  \frac{{\tilde C}_4^A}{M^2} (g_{\mu \nu} q\cdot p' - q_{\nu} p'_{\mu})+
 {{\tilde C}_5^A} g_{\mu \nu}
  + \frac{{\tilde C}_6^A}{M^2} q_{\nu} q_{\mu}\right] \gamma_5
\end{eqnarray}
while for the case of spin half states $\Gamma_{ \mu }$ is generally expressed in terms of 
four form factors as,
\begin{align}  \label{eq:vectorspinhalfcurrent}
  V^{\mu}_{\frac{1}{2}} & = \frac{{F_1}(Q^2)}{(2 M)^2}
  \left( Q^2 \gamma^\mu + \slashchar{q} q^\mu \right) + \frac{F_2(Q^2)}{2 M} 
  i \sigma^{\mu\alpha} q_\alpha \nonumber \\ 
  A^{\mu}_\frac{1}{2} &= - {F_A}(Q^2) \gamma^\mu \gamma^5  - 
  \frac{F_P(Q^2)}{M} q^\mu \gamma^5,
\end{align}
The vector form factors for the resonant states (except for the $\Delta$-resonance) are parameterized 
using helicity amplitudes from the MAID analysis. 
The parameterizations and various form of vector form factors used in the present calculations are given in Ref.~\cite{aip1}.
 For the axial form factors we have used the Goldberger-Trieman relation which 
relates the $R\to N\pi$ coupling to the $\tilde C_5^A(0) (F_A(0))$ for the spin three-half(half) resonances.  
To get $R\to N\pi$ coupling strength, we have used 
partial decay width for the different resonant states following PDG values for the partial decay rates.
 The various properties of the resonances along with 
their couplings are tabulated in Table-\ref{tab:included-resonances}.
 Furthermore, assuming PCAC and pion pole dominance at the weak vertex the pseudoscalar form 
factors $\tilde C_6^A(Q^2)(F_P(Q^2))$ are  related to $\tilde C_5^A(Q^2)(F_A(Q^2))$. 
 We have neglected 
the contribution of $\tilde C_{3,4}^A$ form factors for $D_{13}(1520)$ and $P_{13}(1720)$ resonances.

We have also taken deuteron effect in our calculations 
 by following the prescription of Hernandez et al.~\cite{Hernandez:2010bx} and write
 \begin{equation}\label{de}
  \left(\frac{d\sigma}{dQ^2dW}\right)_{\nu d}=\int d{\bf p}_p^d |\Psi_d({\bf p}_p^d)|^2 
  \frac{M}{E_p^d} \left(\frac{d\sigma}{dQ^2dW}\right)_{\rm{ off~ shell}}.
 \end{equation}
  In the above expression $|\Psi_d|^2=|\Psi_0^d|^2~+~|\Psi_2^d|^2$, where $\Psi_0$ and $\Psi_2$ are the
  deuteron wave functions for the S--state and D--state, respectively and 
  have been taken from the works of Lacombe et al.~\cite{Lacombe:1981eg}.
\section{Results and discussions}\label{results}
\begin{figure}[tbh]
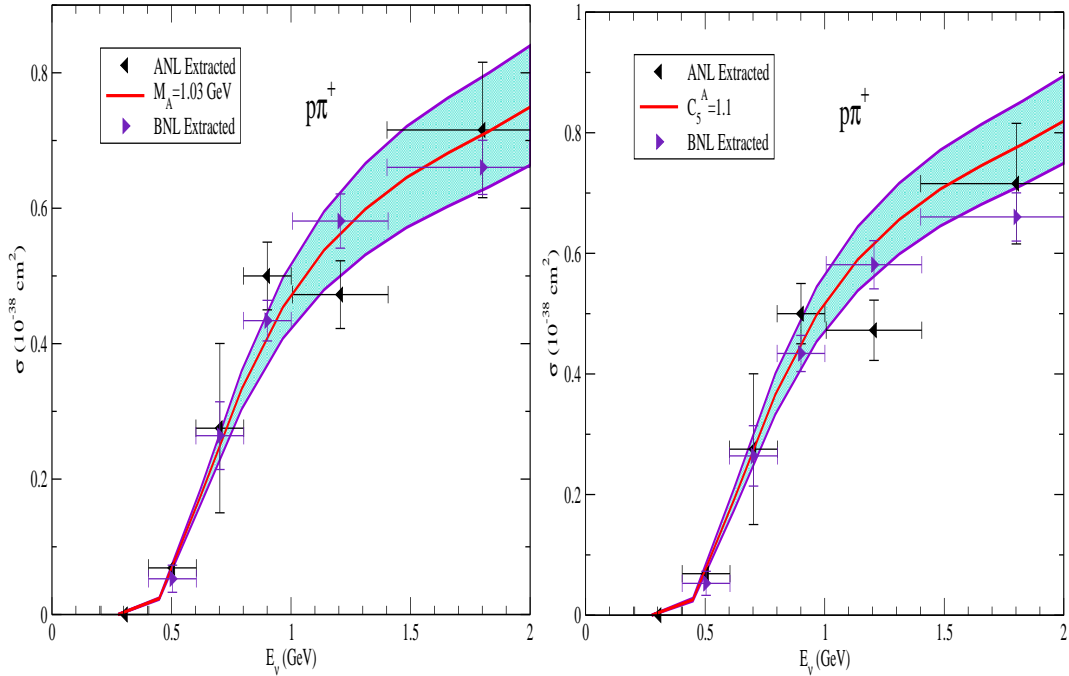

\centering
\includegraphics[height=9cm, width=7 cm]{anl_bnl_ma.eps}
\includegraphics[height=9cm, width=7 cm]{anl_bnl_c5a.eps}
\caption{Total scattering cross section for $\nu_{\mu}   p \rightarrow \mu^{-}   p   \pi^{+}$ process. 
 Data points are reconstructed/reanalyzed data of ANL and BNL experiments by 
 Wilkinson et al.~\cite{Wilkinson:2014yfa}. Here no invariant mass cut has been applied.
In the left panel  change in cross section   with the variation(by 10\%) of axial dipole mass $M_A$ 
 has been shown by taking central value as the world average value. 
While in the right panel the effect of variation of axial charge for $\Delta(1232)$ resonance has been shown.
The central curve has $\tilde C_5^A(0)|_{\Delta} = 1.1$ and the shaded region has been obtained  
by varying $\tilde C_5^A|_{\Delta}$
by 10\%. }
\label{fig1}
\end{figure}

Using the expression for the differential scattering cross section given in Eq.~\ref{eq:phsp}
and integrating over the kinematical variables we obtain the result for total scattering cross section.
To incorporate the deuteron effect we have used Eq.~\ref{de}. In all the numerical calculations where $M_A$ appears, we have taken it as the world average value, 
i.e. $M_{A}=1.026$ GeV.  

In Fig.~\ref{fig1}, we have shown the results for the total scattering cross section for the 
  charged current neutrino induced 1$\pi^+$ production process on proton target 
  i.e. for the reaction $\nu_\mu  p \rightarrow \mu^-  p  \pi^+$.
   The results are presented for the total scattering cross section with 
   $\Delta(1232)$ and non-resonant 
   background(NRB) terms. The results presented 
   here are obtained without using any cut on invariant mass. We have compared the results with the reanalyzed experimental data of
   ANL~\cite{Radecky:1981fn} and BNL~\cite{Kitagaki:1986ct} experiments by Wilkinson et al.~\cite{Wilkinson:2014yfa}. 
 Furthermore, the effect of  varying $\tilde C_{5}^{A}(0)|_\Delta  $ and $M_{A}$ on total  scattering cross section has been studied. 
 We found that the total scattering cross section $\sigma(\nu_\mu  p \rightarrow \mu^-  p  \pi^+)$ 
 has minimum chi-square when  $\tilde C_{5}^{A}(0)|_\Delta = 1.0$ and 
 $M_{A} = 1.026$ GeV are used in the expression of  $\tilde C_{5}^{A}(Q^2)|_\Delta  $.
 However, to see the effects of $\tilde C_{5}^{A}(0)|_\Delta  $ and $M_{A}$ on total scattering cross section 
 we have shown variations of $M_A$ and $\tilde C_{5}^{A}(0)|_\Delta$ in shaded regions.
 We find that the cross section changes by about $ \sim 10\%$ at $E_\nu=1$GeV  if the axial dipole mass $M_{A}$ is varied by 10\%. 
 Similarly, at $E_\nu=1$GeV  when $\tilde C_{5}^{A}(0)|_\Delta $ is varied by 10$\%$ the variation in the cross section is around 9\%.

\begin{figure}[tbh]
\includegraphics[height=9cm, width=15 cm]{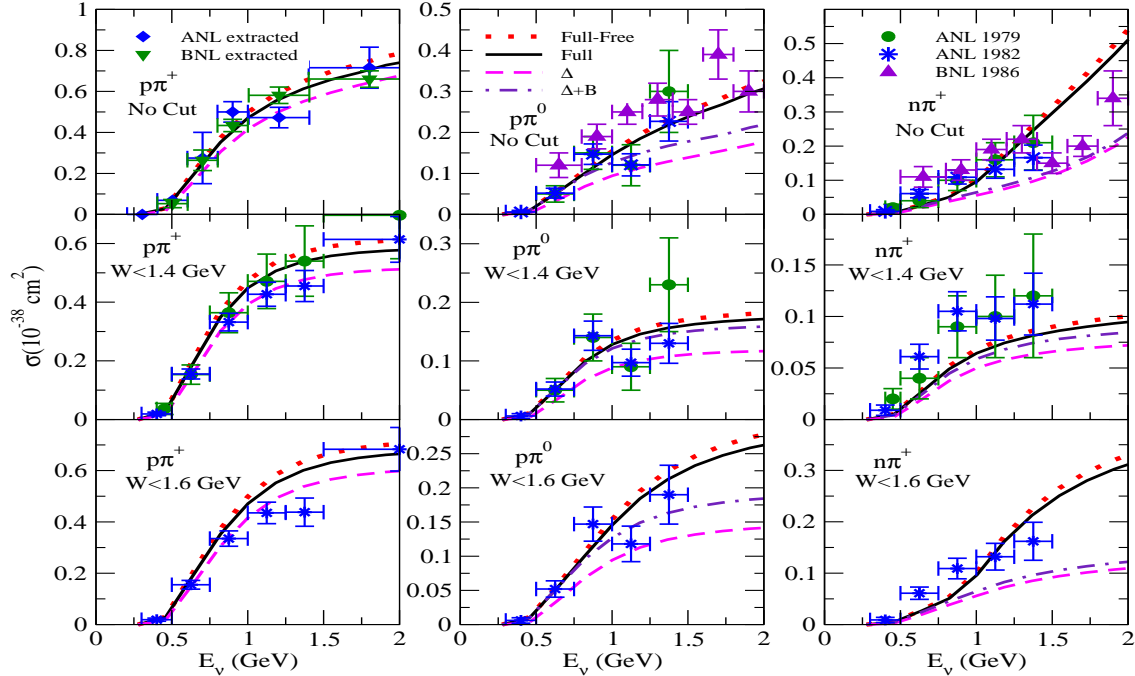}
\caption{Total scattering cross section for the charged current neutrino induced pion production processes through various channels.
 Legends are self explanatory.}
\label{fig2}
\end{figure}
In Fig.~\ref{fig2}, we have presented the results of total scattering cross section for 
the charged current neutrino induced pion production processes in all the channels. 
 The experimental data shown  for $\pi^+ p$ channel is same as in Fig.~\ref{fig1}, while for 
the other channels like $\pi^0 p$ and $\pi^+ n$ the data are from 
ANL~\cite{Radecky:1981fn} and BNL~\cite{Kitagaki:1986ct} experiments.
In the case of $\nu_\mu  p \rightarrow \mu^-  p  \pi^+$ induced reaction,
 the main contribution to the total scattering cross 
 section comes from the $\Delta(1232)$ resonance and there is no contribution 
 from the higher resonances which are considered here. 
 We find that due to the presence of the non-resonant background
 terms there is an increase in the cross section which is about $12\%$ at $E_{\nu_\mu}$=1GeV
   which becomes $\sim 8\%$ at $E_{\nu_\mu}$=2GeV.

  For $\nu_\mu  n \rightarrow \mu^-  n  \pi^+$ as well as $\nu_\mu  n \rightarrow \mu^-  p  \pi^0$
   processes, there are contributions from the non-resonant background 
   terms as well as from the higher resonant terms besides the $\Delta(1232)$  
  resonance. The net contribution to the total pion production due to the presence of the non-resonant 
  background terms in $\nu_\mu  n \rightarrow \mu^-  n  \pi^+$ 
  reaction results in an increase in the cross section 
  of about $12\%$ at $E_{\nu_\mu}$=1GeV 
 which becomes $6\%$ at $E_{\nu_\mu}$=2GeV. When higher resonances are also taken
 into account there is a further increase in the cross section by about 
  $40\%$ at $E_{\nu_\mu}$=1GeV which becomes $55\%$ at $E_{\nu_\mu}$=2GeV. 
  In the case of $\nu_\mu  n \rightarrow \mu^-  p  \pi^0$ process,  
   due to the presence of the background terms the total increase in the cross section is about 
   $26\%$ at $E_{\nu_\mu}$=1GeV and  $18\%$ at $E_{\nu_\mu}$=2GeV
   and due to the presence of higher resonances there is a further 
   increase of about $35\%$ at $E_{\nu_\mu}$=1GeV and  $40\%$ at $E_{\nu_\mu}$=2GeV.
  Thus, we find that the inclusion of higher resonant terms lead to a significant increase in the cross section for 
  $\nu_\mu  n \rightarrow \mu^-  n  \pi^+$ and $\nu_\mu  n \rightarrow \mu^-  p  \pi^0$ processes. 
  Furthermore, it may also be concluded from the above observations that contribution from non-resonant background
  terms decreases with the increase in
 neutrino energy, while the total scattering cross section
  increases when we include higher resonances in our calculations.
  
\begin{figure}[tbh]
\includegraphics[height=9cm, width=15 cm]{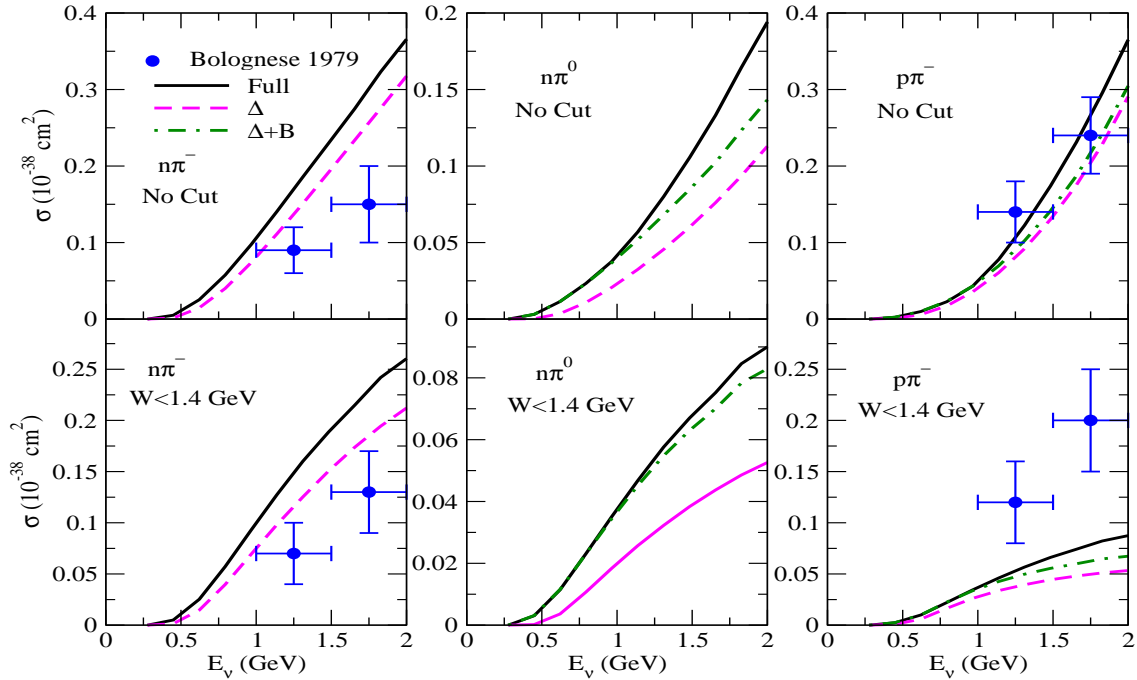}
\caption{Total scattering cross section for the charged current antineutrino 
induced pion production processes through various channels. Legends are self explanatory.}
\label{fig3}
\end{figure}

    When a cut of $W \le 1.4GeV$ on the center of mass energy is applied 
    then due to the presence of the non-resonant background terms,
  the increase in the total scattering cross section
  in the energy range $E_{\nu_\mu}$=1 GeV for  $\nu_\mu  p \rightarrow \mu^-  p  \pi^+$ process   is about $10\%$
  which becomes  $12 \%$ at $E_{\nu_\mu}$=2GeV. 
  For $\nu_\mu  n \rightarrow \mu^-  n  \pi^+$ 
  reaction this increase in the cross section 
  is about $14\%$ at $E_{\nu_\mu}$=1GeV which becomes $5\%$ at $E_{\nu_\mu}$=2GeV.
  When higher resonances are also taken into account there is a further increase in the cross section 
  which is about $40\%$ at $E_{\nu_\mu}$=1GeV which becomes $\sim 55\%$ at $E_{\nu_\mu}$=2GeV. 
  While in the case of $\nu_\mu  n \rightarrow \mu^-  p  \pi^0$   
   due to the presence of the non-resonant background terms the total 
   increase in cross section is about $26\%$ at $E_{\nu_\mu}$=1-2GeV.  Due to the presence of
   other resonances there is a further increase of about 5$\%$ 
   at $E_{\nu_\mu}$=1GeV, contributions of which become $8 \%$ at $E_{\nu_\mu}$=2GeV.  
     
   In Fig.~\ref{fig3}, we have shown the results for the charged current antineutrino induced pion production processes.
   Here also in the case of $\bar\nu_\mu  n \rightarrow \mu^+  n  \pi^-$ reaction there is
   no contribution from the higher resonances other than $\Delta(1232)$ resonance. 
   The inclusion of non-resonant background terms increases the cross section 
   by around $24\%$ at $E_{\nu_\mu}$=1GeV which becomes
   around $12\%$ at $E_{\nu_\mu}$=2GeV. For $\bar\nu_\mu  p \rightarrow \mu^+  n  \pi^0$ reaction, 
   inclusion of non-resonant background terms increases the cross section by around 
   $42\%$ at $E_{\nu_\mu}$=1GeV which becomes 
   $20\%$ at $E_{\nu_\mu}$=2GeV. When higher resonances are included,
   the cross section further increases by 
   $\sim 2\%$ at $E_{\nu_\mu}$=1GeV which becomes $26\%$ at $E_{\nu_\mu}$=2GeV.
   In the case of $\bar\nu_\mu  p \rightarrow \mu^+  p  \pi^-$ reaction, 
   the inclusion of non-resonant background 
   terms increases the cross section by around
   $16\%$ at $E_{\nu_\mu}$=1GeV which becomes $4\%$ at $E_{\nu_\mu}$=2GeV.
   When higher resonances are included the cross section further increases 
   marginally at $E_{\nu_\mu}$=1GeV and $\sim 15\%$ at $E_{\nu_\mu}$=2GeV.

\section{Conclusions}\label{conclusion}
 We have presented the results for charged current  
  one pion production cross section in the energy region of $E_{\nu / \bar \nu} \leq 2 GeV$. 
 Our model consists of contributions from background terms due to non-resonant diagrams, 
 $\Delta(1232)$ resonant term and the contributions from higher resonances. 
 The $\Delta(1232)$-resonance has the dominant contribution but we also need 
 contributions from the non-resonant background terms and 
 the higher resonant terms to describe the experimental data for all the 
 possible channels of single pion production induced by charged 
 current neutrino/antineutrino induced processes. We used 
 $\nu_\mu p \to \mu^- p \pi^+$ channel to fix the axial charge($C^5_A(0) |_\Delta$) and axial dipole mass $M_A$, 
 as there is no other resonance which contributes to this process.
To fix these parameters,  we have used  reanalyzed data of ANL and BNL  and the 
numerical values obtained from our best fit are $M_A=1.026GeV$ and $C^5_A(0)|_\Delta =1.0$.

    When contribution of higher resonances are taken into account, we find that the 
 major contributions to the pion production come from $P_{11}(1440)$ and $D_{13}(1520)$ resonances. 
 The contribution due to non-resonant terms is more important for $\nu  n \rightarrow \nu  p \pi^-$ process 
 and less important for $\bar\nu   p \rightarrow \bar\nu  p  \pi^0$ process.
 
 The present work contributes to the theoretical understanding of the role of background terms and higher resonance terms in neutrino/antineutrino 
 induced one pion production off the nucleon. It would be interesting to apply the present formalism
 to study the nuclear medium effects in the neutrino/antineutrino induced pion production process 
 from nuclear targets in the accelerator experiments 
 being performed in the few GeV energy region.


\begin{thebibliography}{9}
 
 \bibitem{Radecky:1981fn} 
  G.~M.~Radecky {\it et al.},
    Phys.\ Rev.\ D {\bf 25}, 1161 (1982)
  [Erratum-ibid.\ D {\bf 26}, 3297 (1982)];
  S.~J.~Barish {\it et al.},
    Phys.\ Rev.\ D {\bf 16}, 3103 (1977); 
    Phys.\ Rev.\ D {\bf 19}, 2521 (1979).
  
\bibitem{Kitagaki:1986ct}
  T.~Kitagaki {\it et al.},
    Phys.\ Rev.\ D {\bf 34}, 2554 (1986); Phys.\ Rev.\ D {\bf 42}, 1331 (1990). 
  
\bibitem{Wilkinson:2014yfa} 
  C.~Wilkinson {\it et al.},
    Phys.\ Rev.\ D {\bf 90}, 112017 (2014).
  
 \bibitem{aip1} M. R. Alam, M. Sajjad Athar, S. K. Singh and S. Chauhan, arXiv:1509.08622 [hep-ph].
 
\bibitem{Hernandez:2007qq} 
  E.~Hernandez, J.~Nieves and M.~Valverde,
    Phys.\ Rev.\ D {\bf 76}, 033005 (2007).
    
 \bibitem{Hernandez:2010bx} 
  E.~Hernandez {\it et al.},
    Phys.\ Rev.\ D {\bf 81}, 085046 (2010).

\bibitem{Lacombe:1981eg} 
  M.~Lacombe {\it et al.},
    Phys.\ Lett.\ B {\bf 101}, 139 (1981). 
 
 
\end{thebibliography}
\end{document}